\documentclass[%
 aip,
 sd,%
 amsmath,amssymb,
preprint,%
]{revtex4-1}

\usepackage{graphicx}
\usepackage{dcolumn}
\usepackage{bm}

\begin{document}

\preprint{}

\title{Spectral imaging with dual compressed sensing}

\author{Xue-Feng Liu}
\affiliation{Key Laboratory of Electronics and Information Technology for Space Systems, Center for Space Science and Applied Research,
Chinese Academy of Sciences, Beijing 100190, China}
\author{Wen-Kai Yu}
\author{Xu-Ri Yao}
\author{Bin Dai}
\author{Long-Zhen Li}
\affiliation{Key Laboratory of Electronics and Information Technology for Space Systems, Center for Space Science and Applied Research,
Chinese Academy of Sciences, Beijing 100190, China}
\affiliation{University of Chinese Academy of Sciences, Beijing, 100190, China}%
\author{Chao Wang}
\author{Guang-Jie Zhai}
\email{gjzhai@nssc.ac.cn.}
\affiliation{Key Laboratory of Electronics and Information Technology for Space Systems, Center for Space Science and Applied Research,
Chinese Academy of Sciences, Beijing 100190, China}

\date{\today}

\begin{abstract}
We experimentally demonstrated a spectral imaging scheme with dual compressed sensing. With the dimensions of spectral and spatial information both compressed, the spectral image of a colored object can be obtained with only a single point detector. The effect of spatial and spectral modulation numbers on the imaging quality is also analyzed. Our scheme provides a stable, highly consistent approach of spectral imaging.
\end{abstract}

\maketitle

Spectral imaging, which can capture both the spatial information and spectral information of an object, is of great importance in physics and biology as it can give rich evidence in the diagnoses of matter component and structure \cite{Zagonel, Brooksby, Thaler}. For a spectral image, there are three dimensions of information to be measured, two-dimensional spatial information and one-dimensional spectral information. It is obviously not possible to obtain three-dimensional information in one time measurement with current detectors. As an alternative approach, the detection of spatial image or spectrum should to be performed by scanning, which will lead to mechanical movement and reduce the stability of imaging.

In recent years, a sampling theory called compressed sensing (CS) is derived and attracts widely interests \cite{Howland,Zhao,Yu1,Yu2}. With this theory, one can measure a signal with sampling number far less than Nyquist-Shannon theorem demands \cite{Candes1, Candes2, Candes3, Donoho}. Based on CS theory, Baraniuk et al. proposed an imaging approach named single-pixel camera, in which only a single pixel detector is needed to image a two-dimensional object \cite{Duarte1, Baraniuk, Takhar}. This experiment shows that CS makes it possible to reduce the detection dimensions, which is very important in spectral imaging. The CS theory is also applied in spectral imaging \cite{Willett, Duarte2, August}. Arce et al. obtained spectral images with an array detector needless of scanning through CS theory \cite{Arguello1, Arguello2, Arce, Wu}. In this paper, it is shown that by dual compressed sensing, we can reduce the detection dimensions in both spatial space and spectral space, and only a position-fixed single pixel detector is needed to achieve spectral imaging.

Compressed sensing is a measurement theory in which sampling can be performed with number less than unknown in the signals \cite{Candes1, Candes2, Candes3, Donoho}. The applying of CS theory has two conditions. First, the signal $x$ with unknown number of $p$ must be sparse, or sparse under certain basis such as discrete cosine transform or wavelets. Second, a linear measurement on the signal must be performed. That means, we measure the scalar product of the signal and a measurement matrix $A$ with size $q \times p$ ($q < p$):

\begin{align}
y = Ax + e,
\end{align}

in which $e$ is the measurement noise. As the number of measurement result $y$ is only $q$ which is less than the unknown, Eq.~(1) can not be solved directly. However, if the measurement $A$ is a random matrix, the signal $x$ can be reconstructed by solving the following optimization problem:

\begin{align}
\mathop {\min }\limits_x \frac{1}{2}\left\| {y - Ax} \right\|_2^2 + \tau {\left\| x \right\|_1}.
\end{align}

It is proved mathematically that for a signal with sparsity $k$, it can be constructed accurately through $q \ge Ck\log \left( {n/k} \right)$ measurements, in which $C$ is a constant coefficient \cite{Candes3, Donoho}.

\begin{figure}[htb]
\centerline{\includegraphics[width=8cm]{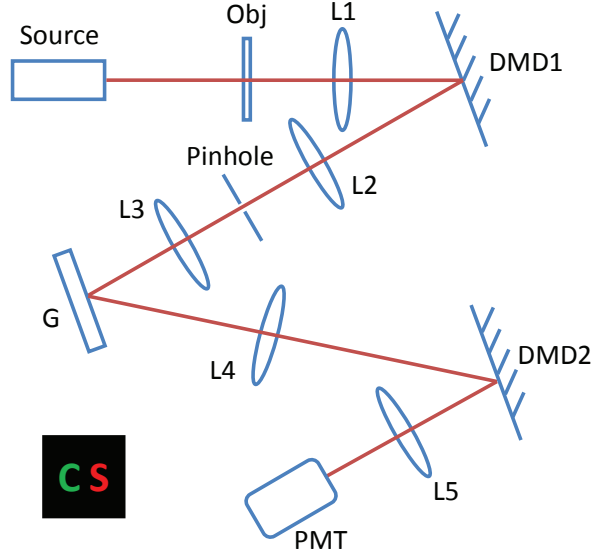}}
\caption{Experimental setup for spectral imaging. Obj, object. G, blazed grating. L1-L5, lens. DMD1, DMD2, Digital micromirror device. PMT, Photomultiplier tube. The DMD reflects the light to two directions. Left bottom, the object to be imaged.}
\end{figure}

The experimental apparatus is given in Fig.~1. A halogen lamp illuminates on the object, which is imaged onto the Digital micromirror device (DMD) by a lens L1. The DMD consists of $1024 \times 768$ micromirrors which can reflect the light to two directions individually. Controlled by the matrix loaded on the DMD1, it can reflect the modulated image to the direction of collecting lens L2, which focuses the light to a pinhole. The light is collimated by the lens L3 and illuminates on the blazed grating. The spectral line will emerge on the focal plane of L4, on which another DMD2 locates. Similarly with DMD1, the DMD2 reflects the modulated spectral line to the lens L5, which collects the light to a photomultiplier tube (PMT).

\begin{figure}[htb]
\centerline{\includegraphics[width=8cm]{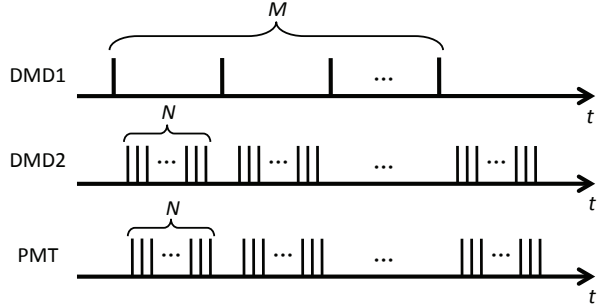}}
\caption{Working timing sequence of DMDs and PMT.}
\end{figure}

The working timing sequence of DMDs and PMT is shown in Fig.~2. $M$ random matrixes $A\left( {x,y} \right)$ are loaded on DMD1 sequentially, modulating the spatial information of the object. When DMD1 states on each modulation, $N$ random matrixes $B\left( \lambda  \right)$ are loaded on DMD2 sequentially, modulating the spectral information of the object. During each modulation of DMD2, the PMT detects the total intensity of object under the spatial modulation $A\left( {x,y} \right)$ and spectral modulation $B\left( \lambda  \right)$.

The spectral image of the object can be expressed by $T\left( \lambda  \right)\left( {x,y} \right)$, in which $\lambda $ denotes the wavelength and $\left( {x,y} \right)$  denotes the spatial coordinate. The modulation process of DMD1 can be described mathematically as

\begin{align}
\sum\limits_{x,y} {A\left( {x,y} \right)T\left( \lambda  \right)\left( {x,y} \right)}  = S\left( \lambda  \right),
\end{align}

in which $S\left( \lambda  \right)$ is the spectrum of the image all over the modulated spatial area, which is also the ¡°image¡± on DMD2. The modulation process of DMD2 can be described as

\begin{align}
\sum\limits_\lambda  {B\left( \lambda  \right)S\left( \lambda  \right)}  = I,
\end{align}

in which $I$ is the total intensity detected by the PMT. Based on CS algorithm, the spectrum $S\left( \lambda  \right)$ can be recovered by $B\left( \lambda  \right)$ and $I$, and then the spectral image $T\left( \lambda  \right)\left( {x,y} \right)$ can be reconstructed by $S\left( \lambda  \right)$ and $A\left( {x,y} \right)$.

In our experiment, the object to be imaged is a film printed a green``C'' and red ``S'' on it, which is shown in the left bottom of Fig. 1. The object is imaged onto $64 \times 64$ pixels of DMD1, and the spectrum line ranging from 520~nm to 620~nm generated on the focal plane of lens L4 occupy $1 \times 1024$ pixels of DMD2. In the spectral imaging of the film, the modulation numbers on DMD1 and DMD2 are $M = 3000$ and $N = 450$, respectively, both of which are sub sampled.

The experimental results are shown in Fig. 3. Based on CS algorithm, in each modulation $A\left( {x,y} \right)$ on DMD1 we can obtain a spectrum line according to the $N = 450$ modulations on DMD2 and correspondent intensity $I$ detected by the PMT, which is shown as a row in Fig.~3(a). Corresponding to spatial modulations on DMD1, $M = 3000$ spectrum lines are recovered. For clear vision, we only show the first 200 spectrum lines in Fig.~3(a). Each column of Fig.~3(a) indicates the intensity fluctuation of the same wavelength. Combined the spatial modulation on DMD1 and intensity fluctuations of different wavelengths, we can reconstruct images of different wavelengths with CS algorithm for the second time.

\begin{figure}[htb]
\centerline{\includegraphics[width=8cm]{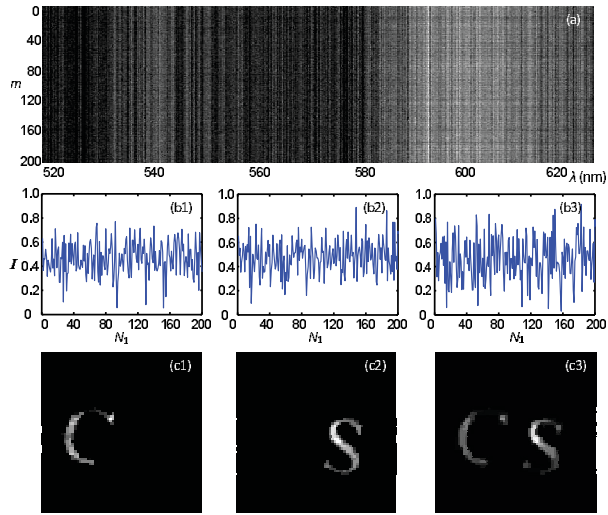}}
\caption{Experimental results of spectral imaging. (a) Spectrum lines recovered. (b1-b3) Intensity fluctuations of different wavelengths, $\lambda = $ 530~nm, $\lambda = $ 610~nm, and all-spectrum. (c1-c3) Imaging results of different wavelengths, $\lambda = $ 530~nm, $\lambda = $ 610~nm, and all-spectrum.}
\end{figure}

In Fig.~3(b1) and (b2), the intensity fluctuations of $\lambda = $ 530~nm and $\lambda = $ 610~nm are shown, respectively. It is obviously that the two curves are different, because the spatial distributions on images of various wavelengths are different, although they correspond to the same modulations of DMD1. The images of the two wavelengths are then reconstructed, giving the green ``C'' and red ``S'', which is shown in Fig.~3(c1) and (c2). From the imaging results, by our spectral imaging system with dual compressed sensing, we can obtain the images of various wavelengths without affected by other spectral components. If we integrate the intensity in each row of Fig.~3(a) over the whole spectrum, the spectral information will be vanished and the image without spectrum can be obtained, similarly to the conventional compressed sensing imaging. The intensity fluctuation of the all-spectrum imaging is shown in Fig.~3(b3) and the corresponding imaging result in Fig.~3(c3), in which a clear image of``CS'' emerges.

The scheme of spectral imaging with dual compressed sensing can achieve spectral imaging with only a single point detector, while the imaging time will be expended by the modulations of the two DMDs. In CS imaging, the dimensions of detector are reduced with sampling numbers increased at the same time. Therefore, there is a tradeoff between the detection dimensions and the imaging time. The dual compressed sensing in our scheme saves the dimensions both in the spatial and spectral detections, and costs more time than conventional CS imaging. The imaging time is decided by the product of numbers of spatial modulation $m$ on DMD1 and spectral modulation $n$ on DMD2.

\begin{figure}[htb]
\centerline{\includegraphics[width=8cm]{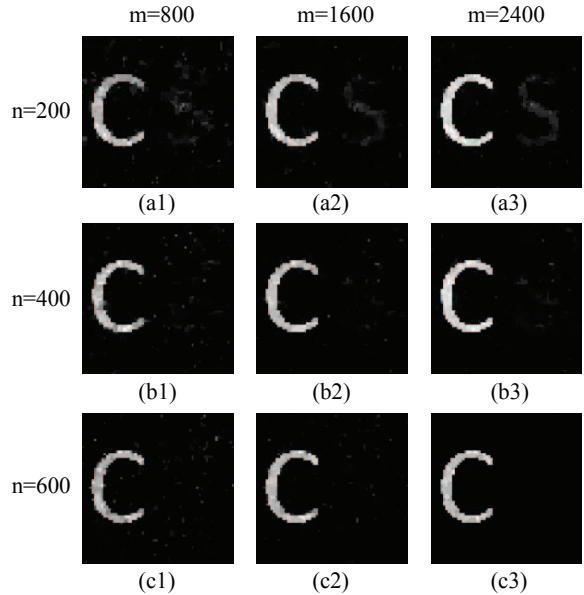}}
\caption{Numerical simulation results of reconstructions for various numbers of spatial modulation $m$ and spectral modulation $n$.}
\end{figure}

To study the impact of sampling numbers on the imaging quality, a simulation is performed. The spectral object is still a green ``C'' and red ``S'', the center transmission wavelengths are 500~nm and 600~nm, respectively, with spectral widths both 40~nm. The image pixel is $64 \times 64$, and the spectrum ranging from 400~nm to 700~nm expands on $1 \times 1024$ pixels. The image with spectrum range of 480~nm to 520~nm is recovered to reveal the green ``C''.

Fig.~4 presents the reconstructions for various numbers of spatial modulation $m$ and spectral modulation $n$. In the three rows, the number of $n$ is 200, 400, and 600, respectively, and in the three columns, the number of $m$ is 800, 1600, and 2400, respectively. It is obviously that the increase of both $m$ and $n$ can improve the imaging quality, while the particular impacted aspects are different. In Fig.~4(a1-a3), the sampling number in spectral information $n$ is only about 20\% of the spectrum pixels, causing inaccurate in the reconstruction of spectrum, resulting in the emergence of ``S'' in the reconstruction results, as the information of other wavelengths is contained in the intensity fluctuations of spectrum used to recover the ``C''. When the spectral sampling number increases, the spectrum can be reconstructed accurately and the red ``S'' will not be imaged, as shown in Fig.~4(b) and Fig.~4(c). From the left column to the right column, the increase of spatial sampling number $m$ can dramatically decrease the noise in the images, while it is unrelated with the spectrum reconstruction accuracy. In Fig.~(a), with the increased $m$, the image of ``S'' even tends to be more clear. As the intensity fluctuations of the reconstructed spectrum have contained the information of ``S'', the increase of spatial samples can improve the quality of ghost image which should not appear with the right image simultaneously. From Fig.~4, it is shown that the spectral modulation $n$ decides the accuracy of spectrum reconstruction, and the spatial modulation $m$ affects the noise level in the image. This conclusion is meaningful in the design of experiment. In fig.~4(a3) and Fig.~4(c1), the sampling number decided by $m \times n$ is the same, while the spectral imaging result is quite different. In fig.~4(a3), the large number of spatial modulations $m$ makes the imaging noise be very low. However, the lack of enough spectral modulations $n$ causes ghost image in the spectral image, which is fatal in most spectral imaging applications, as spectrum information is usually with high interest. In fig.~4(c1), the ghost image is eliminated by enough spectral modulations $n$. Although there is a relatively high noise level, it can be distinguished artificially according to experience in most situations. Therefore, in spectral imaging, the proper choice of spatial modulations $m$ and spectral modulations $n$ is important for a finite sampling time. To obtain accurate spectrum information, enough number of spectral sampling should be guaranteed.

In conclusion, we experimentally demonstrated a spectral imaging scheme with dual compressed sensing. We successfully obtain the image of a spectral object with only a point detector, with spatial sampling and spectral sampling number both compressed. We analyze the effect of spatial modulation number and spectral modulation number on the imaging quality. Through simulation, we find that the spectral modulation number must be enough to avoid emerging false ghost image in the result. As our spectral imaging scheme does not need mechanical movement and each measurement is an overall sampling, it has satisfying stability and consistency. Therefore, we hope that it will have wide applications in material, biology and other fields in which spectral imaging is interested.

This work was supported by the National Major Scientific Instruments Development Project of China (Grant No. 2013YQ030595), the National High Technology Research and Development Program of China (Grant No. 2013AA122902), and the National Natural Science Foundation of China (Grant No. 11275024).


\end{document}